\documentclass[12pt]{article}
\usepackage[cmtip,arrow]{xy}
\usepackage{pb-diagram,pb-xy}
\usepackage{amsmath}
\usepackage[psamsfonts]{amssymb}
\usepackage{cmmib57}
\usepackage{amsmath,amssymb,amscd}
\newcommand{\bPf}{\par\vspace*{-4pt}\indent{\sc Proof.}\enskip}
\newcommand{\ePf}{\medskip}
\def\QED{\hskip0.1em\hfill\null\ \null\nobreak\hfill\kern3pt\vbox{\hrule\hbox
   {\vrule\kern1pt\vbox{\kern1.7pt\hbox{$\scriptscriptstyle{QED}$}
    \kern0.2pt}\kern1pt\vrule}\hrule}}

\def\END{\hskip0.1em\hfill\null\ \null\nobreak\hfill\kern3pt\vbox{\hrule\hbox
   {\vrule\kern1pt\vbox{\kern1.7pt\hbox{$\,\,\,\vspace{5pt}$}
    \kern0.2pt}\kern1pt\vrule}\hrule}}
\newtheorem{theorem}{Theorem}
\newtheorem{lemma}{Lemma}
\newtheorem{corollary}{Corollary}
\newtheorem{proposition}{Proposition}
\newtheorem{remark}{Remark}
\newtheorem{definition}{Definition}
\newtheorem{example}{Example}

\newcommand{\bCd}{\bEq\begin{CD}}
\newcommand{\eCd}{\end{CD}\eEq}
\newcommand{\bcd}{\beq\begin{CD}}
\newcommand{\ecd}{\end{CD}\eeq}
\newcommand{\ben}{\begin{enumerate}}
\newcommand{\een}{\end{enumerate}}
\newcommand{\bEq}{\begin{eqnarray}}
\newcommand{\eEq}{\end{eqnarray}}
\newcommand{\beq}{\begin{eqnarray*}}
\newcommand{\eeq}{\end{eqnarray*}}
\newcommand{\bDf}{\begin{definition}\em}
\newcommand{\eDf}{\end{definition}}
\newcommand{\bLm}{\begin{lemma}}
\newcommand{\eLm}{\end{lemma}}
\newcommand{\bPr}{\begin{proposition}}
\newcommand{\ePr}{\end{proposition}}
\newcommand{\bTh}{\begin{theorem}}
\newcommand{\eTh}{\end{theorem}}
\newcommand{\bCr}{\begin{corollary}}
\newcommand{\eCr}{\end{corollary}}
\newcommand{\bRm}{\begin{remark}\em}
\newcommand{\eRm}{\end{remark}}
\newcommand{\bEx}{\begin{example}\em}
\newcommand{\eEx}{\end{example}}
\newcommand{\C}{\mathbb{C}}

\newcommand{\ie}{{\em i.e$.$} }
\newcommand{\eg}{{\em e.g$.$} }
\newcommand{\R}{I\!\!R}

\newcommand{\cE}{\mathcal{E}}

\newcommand{\cL}{\mathcal{L}}

\newcommand{\bG}{\boldsymbol{G}}

\newcommand{\bK}{\boldsymbol{K}}

\newcommand{\bP}{\boldsymbol{P}}

\newcommand{\bR}{\boldsymbol{R}}

\newcommand{\bX}{\boldsymbol{X}}
\newcommand{\bY}{\boldsymbol{Y}}

\newcommand{\del}{\delta}
\newcommand{\eps}{\epsilon}

\newcommand{\lam}{\lambda}
\newcommand{\sig}{\sigma}

\newcommand{\For}{{\Lambda}}

\newcommand{\Var}{{\mathcal{V}}}
\newcommand{\Thd}{{\Theta}}
\title{\large{\bf Topological obstructions in Lagrangian field theories, with an application to $3$D Chern--Simons gauge theory}}
\author{
{\normalsize Marcella Palese} 
\\ 
{\footnotesize Department of Mathematics, University of Torino}
\\
{\footnotesize via C. Alberto 10, 10123 Torino, Italy}
\\  
{\footnotesize e--mail: {\sc marcella.palese@unito.it
}}
\\ 
{\normalsize Ekkehart Winterroth}
\\
{\footnotesize Department of Mathematics, University of Torino, Italy}
\\ 
{\footnotesize via C. Alberto 10, 10123 Torino, Italy}
\\
{\footnotesize  and  Lepage Research Institute, 17 November 1, 081 16 Pre\v sov,
Slovak Republic}
\\
{\footnotesize e--mail: {\sc  ekkehart.winterroth@unito.it}}
}
\date{}
\begin{document}

\maketitle

\begin{abstract}
We relate the existence of Noether global conserved currents associated with locally variational field equations to existence of global solutions for a local variational problem generating global equations.
Both can be characterized as the vanishing of certain cohomology classes.

In the case of a $3$-dimensional Chern--Simons gauge 
theory, the variationally featured cohomological obstruction to the existence of global
solutions is sharp and equivalent to the usual obstruction in terms of the Chern 
characteristic class for the flatness of a principal connection. 

We suggest a parallelism between the geometric interpretation of characteristic classes as obstruction to the existence of flat principal connections and  the interpretation of certain de Rham cohomology classes to be the obstruction to the existence of global extremals for a local variational principle.
\end{abstract}



\section{Introduction}

As well known, Chern and Simons proved that real cohomology classes arise in the total space of a principal fibered bundle and depend on a connection in that bundle.
We provide, for general field theories on bundles, a (variationally featured) analogue of Chern characteristic classes on principal bundles \cite{ChSi71,ChSi74}.

We obtain a general statement ({\em Theorem \ref{Th1}}) saying that if a variational problem admits global critical sections then all conservation laws derived from symmetries of the field equations admit global conserved quantities.
This result relays on the relation between globality of conserved currents associated with invariance of (locally) variational field equations 
and certain cohomology classes defined by the dynamical forms themselves and their symmetries; such classes are isomorphic to certain de Rham cohomology classes of the space of fields.
We relate such classes to cohomology classes of the basis manifold of the theory by pull-back {\em via} global sections.
If one of such sections defines a non vanishing cohomology class, then there is no global critical section in its homotopy class ({\em Corollary \ref{Cr1}}).
This is strictly based on the `modulo contact' structure of the calculus of variations on fibered manifolds.

In particular, we prove that in the case of a (closed) $3$-dimensional Chern--Simons gauge 
theory, the cohomological obstruction to the existence of global
solutions is sharp and equivalent to the usual obstruction in terms of the Chern 
characteristic class for the flatness of a principal connection ({\em Theorem \ref{Th2}}). 

Chern--Simons field theories have been used, in particular, as a model for classical and quantum gravitational fields \cite{DJT82}.  
In general, just to mention a few, applications range
from classical field theory 
to quantization of Chern--Simons gauge theory of gravity and string theory.
Considering gravity as a gauge theory in all odd dimensions and particularly in dimension three, where the field equations reproduce Einstein field equations, a Chern--Simons Lagrangian can be considered (instead of the Hilbert--Einstein Lagrangian) in which the gauge potential is a linear combination of a frame
and a spin connection;  in particular, $(2+1)$ gravity with a negative cosmological constant can be formulated as a Chern--Simons theory (see, \eg \cite{Wi88,Wi89}, as well as \cite{Ch89,Ch90}  for higher dimensional
Chern--Simons gravity). Developing a $3$-dimensional Chern--Simons theory as a possible and
simpler model to analyse $(2+1)$ gravity  
brought in particular results concerned with  thermodynamics of higher
dimensional black  holes \cite{BaTeZa92}, which in turn produced
a renewed interest in Chern--Simons theories and, consequently, in the
problem of gauge symmetries and gauge charges for Chern--Simons theories \cite{BCDPS13,BFF03,BFFP03,BrKr05,BrKr07,
DJT82,FrPaWi13JPCS,GiMaSa03,PaRoWiMu16}.

In general, from a variational point of view, by adding a Chern--Simons term to the Lagrangian \cite{DJT82}, it appears that `topological contents' are added to the theory and this has of course many potential concrete applications in both classical and quantum theories. Beside theoretical field theory, Chern--Simons terms were proposed also to describe topological aspects involved in statistical mechanics of polymers \cite{FeLa98,FeLa99}. Other approaches in higher dimensions are so-called Lie algebras expansion methods: in  \cite{EdHaTrZa06} gravity as a Chern--Simons action for the AdS algebra in five dimensions is modified through an expansion of the Lie algebra.

Higher dimensional Chern--Simons terms or Chern--Simons type modifications (also called `transgressions') appear in a variety of physical theories, \eg in sigma-model anomaly in brane theory and supergravity.
In particular the globality of solutions of equations of motion is physically relevant under several aspects.
In a recent paper it was proved that, under certain conditions, for certain classes of metrics, adding  Chern--Simons Lagrangian terms does not changes solutions of 
 field equations 
\cite{BCDPS13}; whereby  `the related question of whether the effects of gravitational Chern--Simons terms are of topological nature or not' was mentioned as a `far from obvious question'.

It is therefore especially important  {\em to investigate how the topological nature of the field bundles themselves could be affected}.

In order to study the coupling  of a theory with Chern--Simons terms from a topological point of view it is convenient to have at hand cohomology classes defining obstructions analogous to characteristic classes but deriving from the structure of the theory itself as a variational theory.
Indeed, to some extent, most of the statements in \cite{ChSi71} have a variational analogue which descends from the interplay of  Noether Theorems  \cite{Noe18} and variational cohomology. 

Since Noether seminal paper, it appeared clear that field equations coming from a variational problem could be completely described by symmetries of the corresponding Lagrangian and associated conservation laws. 
In particular, from the beginning of the past century globality of field equations has been assumed as a basic requirement for a physical theory to make sense.  
From a foundational point of view it is also quite important that also fields (\ie solutions of field equations) be well defined intrinsic objects and globally defined.
Sheaf theory and cohomology are indeed the right mathematical tools to deal with such problems.

We restrict our considerations to  {\em global field equations} which are  {\em (locally) variational}, this means that, at least locally, they can be derived variationally from a Lagrangian, which is then defined on an open set of the relevant field space. In particular, local Lagrangians differ  by a local variationally trivial Lagrangian (a local divergence). Specifically, Chern--Simons gauge theories are of this type.

Noether already pointed out the cohomological content of the invariance of the Euler-Lagrange expressions and dealt with the study of the relationship between currents associated with symmetries of {\em global} Lagrangians and of corresponding Euler-Lagrange expressions; an aspect further specified by Bessel-Hagen \cite{BeHa21}.
It is now of a certain importance (for the study of cohomological obstructions to the existence of global solutions) that, for {\em local} presentations of a variational problem, the corresponding conserved currents are, in principle, {\em local currents}, see \eg
\cite{FePaWi10,FrPaWi13,FrPaWi13JPCS,PaRoWiMu16}. 
However, symmetries of such {\em local Lagrangians} are also symmetries of the associated {\em global Euler-Lagrange equations}.
Thus locally defined canonical Noether currents are associated with global symmetries of global field equations. A somehow apparently paradoxical feature! We also recall that, in general, not all symmetries of the Euler-Lagrange equations are also symmetries of the corresponding variational problem, and therefore not all of them generate Noether currents \cite{CaPaWi16}. At a first glance, these features could appear as subtleties, however, we shall see that they are fundamental to distinguish what can be provided with a physical meaning. 

There exists a cohomological obstruction for Noether--Bessel-Hagen local currents be globalized \cite{FePaWi10,FrPaWi13JPCS} 
and we prove that this is also an obstruction to the existence of global critical sections. 

It is important to stress that Noether--Bessel-Hagen currents are defined as 
{\em the difference between canonical Noether currents -- originating from the invariance of a Lagrangian -- and 
 Bessel-Hagen currents -- originating from invariance of equations}, see below; we stress that the interplay of these two invariances (and of their {\em independent} locality/globality characters) is the object of this research and {\em not} simply the study of Bessel-Hagen, also-called generalized or `modulo divergence' Lagrangian symmetries.

In the case of a purely gauge Chern--Simons theory in dimension three, we find that such a variational obstruction is sharp \ie it is equivalent to the usual obstruction given by the first Chern characteristic class. The proof of this equivalence is an original result and did not appear before in the literature.

\section{From local to global in the calculus of variations}

To investigate obstructions for local variational objects to be globalized, we shall use a modern geometric formulation of the calculus of variations as a subsequence of the de Rham sequence of differential forms on finite order prolongations of fibered manifolds (in most applications in theoretical physics these are chosen to be gauge(-natural) bundles).

Let us shortly  summarize the geometric frame. 
Lepage for first showed that within the multiple integrals formulation of the calculus of variations, which is at the basis of classical formulation of field theory, a Lagrangian could be obtained as a differential form modulo a suitable contact structure (\ie in his words through a {\em congruence} \cite{Lep36}).
This fundamental idea lead  to geometric constructions known as {\em variational complexes} (for a review and several references, see \eg \cite{Kru08,Kru15,PaRoWiMu16}) which are in fact the contemporary formulations enabling to go from local to global in the calculus of variations. The main point already {\em in nuce} in the work by Lepage is that the cohomology of a variational sequence is isomorphic to the de Rham cohomology of the field space. We shall then consider {\em variational forms}, \ie equivalence classes of differential forms modulo a certain contact structure. 
We geometrize field and  field equations by means of jets (equivalence classes of maps between the manifold of `space' coordinate and the manifold of  `fields' having the same Taylor expansion up to a certain order $r$). Such objects can be geometrized by a $r$-jet bundle. The `dynamical' content of equations is therefore completely encoded by the contact structure related to the natural projection from any order $r$ to the immediately subsequent order $r-1$.

We assume the $r$-th order prolongation of a fibered manifold $\pi: \bY \to \bX$, with $\dim \bX = n$ and $\dim \bY = n+m$, to be the configuration space; this means that {\em fields are assumed to be (local) sections} of $\pi^{r}: J_r \bY \to \bX$.
Due to the affine bundle structure of  
$\pi^{r+1}_{r}: J_{r+1} \bY \to J_{r} \bY$, we have a natural splitting
$J_r\bY\times_{J_{r-1}\bY}T^*J_{r-1}\bY =
J_r\bY\times_{J_{r-1}\bY}(T^*\bX \oplus V^*J_{r-1}\bY)$, which induces natural  {\em splittings} in  horizontal and vertical parts of vector fields, forms and of the exterior differential on $J_r\bY$.
Starting from this splitting one can define sheaves of contact forms $\Thd^{*}_{r}$ defined by the kernel of the horizontalization. The local generators of the contact ideal are well known pfaffian $1$-forms defining higher order partial derivatives, and their differentials.
The sheaves $\Thd^{*}_{r}$ form an exact subsequence of the de Rham sequence on $J_r\bY$ and one can define the quotient sequence
$\{ 0\to \R_{\bY} \to \For^{*}_r/\Thd^{*}_r \,, \cE_{*} \}$, 
called the $r$--th order {\em variational sequence} on $\bY\to\bX$, which is an acyclic sheaf resolution of the constant sheaf $\R_{\bY}$; see \cite{Kru90}. In fact, the cohomology of the complex of global sections $H^{*}_{VS}(\bY)$ is naturally isomorphic to both  the \v Cech cohomology and the de Rham cohomology $H^{*}_{dR}(\bY)$ \cite{Kru90}.
Thus if the cohomology of $\bY$ is trivial, of course each local inverse problem is also global.

The quotient sheaves in the variational sequence can be represented as sheaves $\Var^{k}_{r}$ of $k$-forms on jet spaces of higher order, see \eg \cite{Kru90,PaRoWiMu16}.
Lagrangians are sheaf sections  $\lam\in \Var^{n}_{r}$, while $\cE_n$ is called the Euler-Lagrange morphism; the latter is thus characterized as a quotient morphism of the exterior differential morphism of the de Rham complex; rooting in the foundamental E. Cartan's work \cite{Cartan22} and stemming from Lepage's seminal papers, this idea has been formulated by Dedecker \cite{Ded55,Ded56,Ded57,Ded77} in the language of sheaves and then developed in different aspects by various authors, see \eg 
\cite{AnDu80,Bau,DT80,Kru90,Kru15,Tul77,Tu80,Tu88,Tak79,Vin77}. 

The Euler-Lagrange equations  are therefore given by  $\cE_{n}(\lam)\circ j_{2r+1}\sig =0$ for (local) sections $\sig: \bX \to \bY$.
Sections $\eta\in\Var^{n+1}_{r}$ are called {\em source forms} or  also {\em dynamical forms}, 
while $\cE_{n+1}$ is called the Helmholtz morphism. 

In the case of a nontrivial cohomology of $\bY$, given a closed section of a quotient sheaf of the variational sequence, one look at the problem as to when this section is also globally exact. 
 
 To answer to this question, let $\bK^{p}_{r}\doteq \textstyle{Ker}\,\cE_{p}$; we have a natural short exact sequence of sheaves
which  gives rise in a standard way to a long exact
sequence in \v Cech cohomology,
where the {\em connecting homomorphism}, in this case given by  $\delta_{p} = i^{-1}\circ\mathfrak{d}\circ\mathcal{E}_{p}^{-1}$, is the mapping of cohomologies in the corresponding diagram of cochain complexes (here $\mathfrak{d}$ is the usual {\em coboundary operator}). 

Every global section $\eta\in\cE_{p}(\Var^{p}_{r})$, \ie locally variational, defines a cohomology class 
$\del_{p} (\eta) \in
H^{1}(\bY, \bK^{p}_{r}) $ $\simeq$ $ H^{p+1}_{VS}(\bY)
$ $\simeq$ $ H^{p+1}_{dR}(\bY)$. Every non vanishing cohomology class in  $H^{p}_{dR}(\bY)$ gives rise to local variational problems.
It is clear that $\eta$ is globally variational if and only if $\del_{p} (\eta) = 0$. 

\subsection{A fundamental cohomology class associated with field equations}

We shall explicate now how cohomology enters in globality problems concerned with conserved quantities. We shall examine this aspect within Noether formalism \cite{Noe18}. 

Factorizing modulo contact structures 
 is the basic idea underlying the definition of  a 
{\em variational Lie derivative} operator $\cL_{j_{r}\Xi}$ and of a {\em quotient Cartan formula} (a {\em variation  formula}) defined  on the sheaves of the variational sequence. This enable us to define symmetries of classes of forms of any degree in the variational sequence and corresponding  conservation theorems; see \cite{CaPaWi16,PaRoWiMu16} for details.

The Cartan formula for the variational Lie derivative of closed classes of forms (exterior differential forms modulo the contact structure induced by affine projections $\pi^{r+1}_{r}$ projections) selects a quite important class defined by both the vertical part of the symmetry and the Euler-Lagrange class.

In fact  we have that $\del_p (\cL_{\Xi}\eta_{\lam}) = 0$, \ie the variational Lie derivative `trivializes' cohomology classes.
By the way, although a fact not in the core of the present paper, it is  remarkable that, if the first variational derivative of a local presentation is closed, therefore the second variational derivative define a trivial cohomology class. 
In fact, the first variational derivative -- with respect to symmetries of Euler-Lagrange expressions --  of a local presentation is closed; as a consequence a trivial cohomology class is defined by the variational derivative of `strong' Noether currents \cite{FePaWi10,FrPaWi13}.

Let then now $\eta_{\lam}$ denotes a global Euler--Lagrange morphism for a local variational problem $\lam_i$ (in the following a subscript `i' indicates that we are dealing with a local object defined on a open set $U_i$ in a good covering); if $\eps_i$ denotes a local canonical Noether current, for a projectable vector field $\Xi$, Noether's First Theorem reads {\em locally} 
$
\cL_{\Xi} \lam_{i}= \Xi_V\rfloor\eta_\lam +d_H \eps_i
$. 

Since we assume $\eta_\lam$ to be closed, the quotient Cartan formula reduces to $\cL_{\Xi} \eta_\lam =\cE_n (\Xi_V\rfloor \eta_\lam)$, and  if $\Xi$ is such that $\cL_{\Xi} \eta_\lam =0$, then $\cE_n (\Xi_V\rfloor \eta_\lam) = 0$;
therefore {\em locally} 
$\Xi_V\rfloor\eta_\lam =d_H\nu_i$.
Notice that $\mathfrak{d}(\Xi_V\rfloor \eta_{\lam}) = 0$, but in general we have the obstruction 
$\del_{n-1}(\Xi_V\rfloor \eta_{\lam})\neq 0$, 
so that the current  $\nu_i$ is a {\em local} object (conserved along the solutions of Euler--Lagrange equations \ie critical sections).
On the other hand, {\em and  independently} (see \cite{Noe18,BeHa21}), for a symmetry of $\eta$ but not of $\lam$, we get {\em locally}
\beq
\cL_{\Xi} \lam_{i}=d_H \beta_{i}\,,
\eeq
thus  we can write
$\Xi_{V} \rfloor \eta_{\lam}  + d_{H}( \eps_i  -  \beta_{i} )$  $=$ $0$. 
\bDf
We call the (local) current $\eps_{i} - \beta_{i}$ a {\em Noether--Bessel-Hagen current}. 
\eDf

\bRm
It is important to stress that a Noether--Bessel-Hagen current is the difference between the canonical Noether current -- originating from the invariance of a Lagrangian -- and 
the Bessel-Hagen current -- originating from invariance of equations. In general, one should realize that {\em the obstruction to the globality of the single type of current is given by different and, in principle, indipendent cohomolgy classes}.
\eRm

\bPr
A local Noether--Bessel-Hagen current can be globalized if and only if
\beq 
0 =  \del_{n-1}(\Xi_V\rfloor \eta_{\lam})
\simeq
[\Xi_{V} \rfloor \eta_\lam]  \in H^{n}_{dR}(\bY)\,. 
\eeq 
\ePr
Here $\simeq$ denotes the natural isomorphism between cohomologies.

Note also that, in general, a Noether--Bessel-Hagen current differs from the local potential of the `work' term, here $\nu_i$, by a closed form and  that it is conserved  `on shell'.

\section{A cohomological obstruction to the existence of (global) critical sections}

In the present section we study the cohomology class $[\Xi_{V} \rfloor \eta]$  more closely.
It turns out to be not only the obstruction to the existence of global conserved quantities, but also an obstruction to the existence of global critical sections. 
We start with proving in full detail a result shortly announced in \cite{FrPaWi13JPCS} and an important consequence of it.

\bTh \label{Th1}
Let $\eta$  be the dynamical form of a local variational problem and let   $H^{n}_{dR}(\bY) \sim  \pi^{*} (H^{n}_{dR}(\bX))$. If the variational problem admits (global) critical sections then all conservation laws derived from symmetries of the global field equations admit global conserved quantities.
\eTh
\bPf
The theorem will be proved if we show that the cohomology class $[\Xi_{V} \rfloor \eta]$ vanishes in  $H^{n}(\bY)$. For the sake of simplicity we will omit the specific order of the jets. The variational sequence being a quotient of the de Rham complex, we  know that there exists a closed $n$-form  $\beta$ on (some jet prolongation of) $\bY$, which projects onto $[\Xi_{V} \rfloor \eta]$ and, thus, we have $[\beta] \sim [\Xi_{V} \rfloor \eta]$ in cohomology.
This means, we can write $\beta$ as a sum $\Xi_{V} \rfloor \eta + \alpha + d\gamma$ with $\alpha$ and $\gamma$ contact forms.

Consequently, $\beta$ vanishes along the (jet prolongation of) a critical section  $\sigma$, since the vanishing of $\Xi_{V} \rfloor \eta$ {\em defines} critical sections and  $\alpha$ and $d\gamma$ vanish along {\em all} sections, $\alpha$ and $\gamma$ being contact forms. 
 We can then conclude that
$j \sig^{*}([ \Xi_{V} \rfloor \eta])$ vanishes in $H^{n}(\bX)$.

But by the very definition of (jet prolongation of a) section we know that $j \sig^{*}$ induces an inverse isomorphism to $\pi^*$ in the $n$th cohomology group, \ie $j\sigma^{*} \circ \pi^{*} = 1_{H^{n}_{dR}(\bX)}$ and $\pi^{*} \circ j\sigma^{*} =1_{H^{n}_{dR}(J\bY)}$ on some jet prolongation $J\bY$ of  $\bY$.
Hence, not only  $j \sig^{*}([ \Xi_{V} \rfloor \eta])$ vanishes in $H^{n}(\bX)$, but also $[ \Xi_{V} \rfloor \eta]$ vanishes in  $H^{n}_{dR}(J\bY) \sim H^{n}(\bY)$.
\ePf

\bRm
For practical purposes, the condition  $H^{n}_{dR}(\bY) \sim  \pi^{*} (H^{n}_{dR}(\bX))$ is not very limiting. It is satisfied, for example, by all theories on vector or affines bundle, since in these cases $\bX$ is a contraction of $\bY$; in particular by all theories which can be formulated on the bundle of connections (see below), like Yang-Mills type or Chern-Simons type theories.
\eRm
If, conversely, the class $j \sig^{*}([ \Xi_{V} \rfloor \eta])$ does not vanish, of course, neither $\sigma$ nor any section homotopical to it, \ie any deformation or variation of $\sigma$, can be critical. Thus, $j \sig^{*}([ \Xi_{V} \rfloor \eta])$ is an obstruction to the existence of global solutions for the (local) variational problem defined by $\eta$. 

If $\pi$ and, thus, $j \sig$ induce isomorphisms between  $H^{n}_{dR}(\bX)$ and  $H^{n}_{dR}(J\bY)$, the class $j \sig^{*}([ \Xi_{V} \rfloor \eta])$ vanishes, by the above reasoning,  if and only if $[\Xi_{V} \rfloor \eta]$ vanishes. 
\\ In consequence, we have
\bCr \label{Cr1}
Let $\sigma$ be a section of $\bY$ over $\bX$. 
If $0 \neq j \sig^{*}([ \Xi_{V} \rfloor \eta]) \in H^{n}(\bX)$, then there is no (global) critical section in the homotopy class of $\sigma$. 

Under the condition $H^{n}_{dR}(\bY) \sim  \pi^{*} (H^{n}_{dR}(\bX))$, as in the previous theorem, there are no (global) critical sections if $[ \Xi_{V} \rfloor \eta] \neq 0$.
\eCr

The question `How sharp is this obstruction?' poses itself immediately. Now, Euler-Lagrange equations are essentially a set of partial differential equations and for partial differential equations even the existence of local solutions is a non trivial problem. So, one would think that topological obstructions will hardly be particularly important. However,  we will show in the next section, as a first test case, that the obstruction is sharp for $3D$ Chern--Simons theory. This is an original result and does not seem to have been studied before.

\subsection{The case of $3D$ Chern--Simons gauge theory}

Classical $3$-dimensional Chern--Simons theory is a classical field theory for principal connections on an arbitrary principal bundle $P$ over a $3$-dimensional manifold $\bX$.
It is well known that
the equations can be written $F_{A} = 0$ ($F_{A}$ the curvature of the connection $A$);
Since we are in dimension $3$, the existence of solutions to these equations, i.e flat connections, corresponds to the vanishing of the cohomology class $[tr F_{A}]$ (`first Chern class'); recall that this cohomology class is independent of the connection $A$. To simplify our presentation somewhat, we will consider only principal bundles $P$ with group $U(1)$ or $\C^*$ in the following. This leads essentially to no loss of generality, since we can pass from any principal bundle $\bP$' to a  $U(1)-$ or $\C^*-$principal bundle with the same `first Chern class' by `taking the determinant' in a suitable way.

To apply our result, we have to formulate Chern--Simons theory in terms of  the bundle of connections:
 the bundle of connections is defined as 
 \beq
 \mathcal{C} := J^{1}P/G \mapsto \bP/\bG \sim \bX\,,
 \eeq
 this bundle $\pi_{\mathcal C}: \mathcal{C} \mapsto \bX$ is an affine bundle modeled on the vector bundle $T^* \bX \mapsto \bX$ and, thus, $T{\mathcal C} \sim \pi_{\mathcal C}^* (T^* \bX \oplus T\bX)$, where $\pi_{\mathcal C}^* (T^* \bX)$ corresponds to the vertical part. Note also that $H^{n}_{dR}({\mathcal C}) \sim  \pi^{*} (H^{n}_{dR}(\bX))$ as required in the previous section.
 
 The principal connections on $P$ are in one to one correspondence with the (global) sections of  the bundle $\pi_{\mathcal C}: \mathcal{C} \mapsto \bX$ in the following sense: the contact structure of $J^{1}\bP$ defines a connection on  the principal bundle  $J^{1}\bP \mapsto \mathcal{C}$ (its curvature is given by $\mathcal{F} = dA_{j} \wedge dx^{j}$ for structure groups $U(1)$ or $\C^*$ and fibered coordinates $(x^j, A_j)$ on $\mathcal{C}$) and this connection is universal in the sense that every principal connection on $\bP \mapsto \bX$ is `induced' by it via a section $\sigma: \bX \mapsto \mathcal{C}$, in particular $F_{\sigma} = \sigma^{*}\mathcal{F}$.
To prove the sharpness of the obstruction we will need below the following rather curious observation.
\bPr For every differential from $\gamma$ on $\bX$ there is a vertical vector field $\Xi$ on $\mathcal{C}$ such that $\pi_{\mathcal C}^* (\gamma) = \Xi \rfloor \mathcal{F}$.
\ePr
\bPf
Let $\gamma$ be a differential from on $\bX$ with coordinate expression $\gamma = f_{i} dx^{i}$ and let $\Xi$ be the vertical vector field on $\mathcal{C}$ for which $\Xi =  f_{i} \delta_{A_{i}}$ in coordinates. The proposition follows then at once from the above expression for  $\mathcal{F}$. The global existence of such a $\Xi$ is a consequence of the fact, that  $\pi_{\mathcal C}: \mathcal{C} \mapsto \bX$ is an affine bundle and the vertical part of its tangent bundle can be identified with $\pi_{\mathcal C}^* (T^* \bX)$. In particular, every vertical vector field corresponds to a horizontal form.
\ePf

Chern--Simons theory can be formulated as a variational theory on $J^{1}\mathcal{C}$ starting from Chern--Weil theory and using the construction of secondary characteristic classes by Chern and Simons.
Dealing with the pure gauge theory, it is advantageous to lift the dynamical form of the problem to the de Rham complex:
the pullback from  $\mathcal{C}$ to  $J^{1}\mathcal{C}$ of the second Chern form $\Omega_{2}(\mathcal{F})=\mathcal{F} \wedge \mathcal{F}$ projects onto the Chern-Simons dynamical form $\eta_{CS}$ in the variational sequence. Thus, a section $\sigma$ of 
$\mathcal{C} \mapsto \bX$ is critical if and only if 
\beq
j^{1}\sigma^{*} (\Xi \rfloor (\pi_{0}^{1})^{*} (\Omega_{2}(\mathcal{F})))=j^{1}\sigma^{*} (\Xi \rfloor (\pi_{0}^{1})^{*} (\mathcal{F} \wedge \mathcal{F}))=0\,, 
\eeq
for all vector
fields $\Xi$ on  $J^{1}\mathcal{C}$. The form $\Xi \rfloor (\pi_{0}^{1})^{*} (\Omega_{2}(\mathcal{F}))$ is closed and defines a cohomology class $[\Xi \rfloor (\pi_{0}^{1})^{*} (\Omega_{2}(\mathcal{F}))]$ if and only if $\Xi$ is a Lie symmetry of $(\pi_{0}^{1})^{*} (\Omega_{2}(\mathcal{F}))$, since $(\pi_{0}^{1})^{*} (\Omega_{2}(\mathcal{F}))$ is closed by Chern--Weil theory. This cohomology class is the de Rham counterpart of the class $[ \Xi_{V} \rfloor \eta_{CS}]$ of the previous section.   
 
\bTh \label{Th2}
Chern--Simons theories  on closed, \ie compact without boundary,  $3$-manifolds admit global critical sections if and only if 
\beq
[\Xi \rfloor (\pi_{0}^{1})^{*} (\Omega_{2}(\mathcal{F}))]= 0\,.
\eeq
\eTh

\bPf
If there are global critical sections, then $[\mathcal{F}] = 0$, \ie $\mathcal{F} = d\beta$ for some differential form $\beta$ on ${\mathcal C}$, and, thus,  
\beq
[\Xi \rfloor (\pi_{0}^{1})^{*} (\Omega_{2}(\mathcal{F}))]= [2(\pi_{0}^{1})^{*} ((\Xi \rfloor \mathcal{F}) \wedge d\gamma)] =  [2d((\pi_{0}^{1})^{*} (\Xi \rfloor \mathcal{F} \wedge \gamma)] = 0\,.
\eeq
If, conversely, there are no global critical sections, we have $[\mathcal{F}] \neq 0$ and the same holds, by Chern--Weil theory, for every pullback $\sigma^{*}\mathcal{F}$ with a section $\sig: \bX \mapsto {\mathcal C}$, \ie a principal connection on $\bP \mapsto \bX$.

On closed manifolds, exists then, by Poincar\'e duality,  a closed cohomologically non trivial $1$-form 
$\gamma$  such that $\gamma \wedge \sigma^{*}\mathcal{F}$ and $\pi_{\mathcal C}^* (\gamma)  \wedge \mathcal{F}$ are cohomologically non trivial.
By the above proposition there is a vertical vector field $\Xi$ such that $\pi_{\mathcal C}^* (\gamma) = \Xi \rfloor \mathcal{F}$ and, thus, 
\beq
[\Xi \rfloor (\pi_{0}^{1})^{*} (\Omega_{2}(\mathcal{F}))] = [\Xi \rfloor (\pi_{0}^{1})^{*} (\mathcal{F} \wedge \mathcal{F})] = [2(\pi_{0}^{1})^{*} (\pi_{\mathcal C}^* (\gamma)  \wedge \mathcal{F}) \neq 0\,.
\eeq
This concludes the proof.
\ePf

For non compact manifolds there is no Poincar\'e duality and the result is, of course, no longer true. However, if the model can be extended onto a `suitable' compactification the result can still be applied. `Suitable' means that all mathematical objects representing physical quantities can be extended (\eg a metric with signature $+--$ can be extended from $\bR^3$ to $S^1 \times S^2$ but not to $S^3$). Physically, such a compactification can be interpreted as the requirement that all physical data be finite.

\section{Conclusions and perspective}

Under certain (not much restrictive) conditions, we have shown that if a variational problem admits (global) critical sections then all conservation laws derived from symmetries of the field equations admit global conserved quantities.
Furthermore, we have identified an obstruction to the existence of global solutions for the variational  problem defined by a given dynamical form. 
In particular, we have shown that there is no (global) critical section in the homotopy class of each global section defining a non vanishing cohomology class. 

In the case of a purely gauge Chern--Simons theory in dimension three, we have found that our variational obstruction is sharp \ie it is equivalent to the usual obstruction given by the first Chern characteristic class. 

The natural continuation is now to 
study our variational cohomology class $[\Xi_{V} \rfloor \eta]$ in higher dimensional Chern--Simons theories \cite{Win16}.
These theories are used for describing the quantum Hall effect and the conditions under which they admit global solutions seem not to be stated in the literature. 

Chern--Simons terms in theories of topological gravity proposed in the existing  wide literature appear quite different among them, all of them sharing the characteristic of being some kind of `transgression' for some kind of `curvature', the underlying geometric structure (principal bundle structure) remaining sometimes unexploited; the gauge-natural nature of prolongations of principal connections remaining aside \cite{FFPW08,FFPW11}.
It seems of interest in those cases, depending on the effective choice of the form of Chern--Simons terms, to investigate the obstruction for the existence of global critical sections (\ie of global Noether--Bessel-Hagen currents). 

In order to obtain new global Noether--Bessel-Hagen currents, additional (local) transgression terms indeed should be always chosen in such a way that there exist global critical sections for the whole variational problem.  From this point of view, according to \cite{BCDPS13}, it is relevant to understand if and how solutions of equations of motions are affected by adding such terms.
Of course, the vanishing of added dynamical terms (some kind of `curvature') along solutions grants that the corresponding obstruction to globality of conserved quantities vanishes. This is \eg the case of the Cotton tensor in topologically massive gauge theories which vanishes along Einstein solutions. 
The variational obstruction, of course, obviously vanishes if the `curvature' terms added to the equations of motions are identically vanishing (\ie along any sections).
The latter two cases are sufficient, but not necessary conditions for possible local modifications of the original Lagrangian to generate global currents associated with the modified (anyway global) equations. The necessary condition is ultimately given by the vanishing of the cohomology class defined by the closed form obtained by contracting the dynamical form with symmetry generators.  Such a condition is satisfied when this contraction is globally exact. Of course there are classes of theories satisfying it (\eg all gauge-natural theories, canonically, along the kernel of the Jacobi morphism \cite{PaWi07}).

Indeed, the r\^ole of the Chern--Simons theory is well recognized in many aspects of physics and natural sciences whenever a symmetry group  is underlying, and a principal bundle structure can be recognized as fundamental for the definition of infinitesimal transformations of fields. In many cases, symmetry of field equations are not known in advance and have to be determined.
Our approach could shed in particular a new light also on zero-curvature type approaches to integrable systems, and, in a more abstract context, on the integrable towers formulation of nonlinear systems, see \eg  \cite{PaWi02,PaWi12}.
In fact, the obstruction defined by the contraction of the dynamical form with generators should determine either the significant additional terms, as mentioned before, or, which is the case in the latter models, once fixed the form of the added term, the meaningful symmetry transformations.

The vanishing of this `variational cohomology class' itself could be interpreted somehow as a flatness condition for an underlying principal bundle structure for any variational problem on general bundles. 
This aspect is currently under investigation and will be the subject of a separated paper. 

As a final remark, it should also be stressed  that Jacobi forms \cite{PaWi07,PaWi11} (related to linearized equations of motion) contracted with generators of symmetries define in general a different cohomology class, so that the cohomology class defined by the contraction of $\eta$ could be trivial, but not the one defined by  the contraction of the variation of $\eta$.

\section*{Acknowledgements} 
Paper partially supported by MIUR {\em via} Department of Mathematics - University of Torino under research project {\em  Strutture geometriche e algebriche in fisica matematica ed applicazioni -
 2015} (MP scientific responsible), GNSAGA-INdAM and Lepage Research Institute.



\begin{thebibliography}{99}

\bibitem{AnDu80} I. M. Anderson, T. Duchamp: On the existence of global variational principles,
{\em Amer. Math. J.} {\bf 102}
(1980) 781--868.

\bibitem{BaTeZa92}  M. Ba\~{n}ados, C. Teitelboim, J. Zanelli:
Black hole in three-dimensional spacetime, {\em  Phys. Rev. Lett.} {\bf  69} (13) (1992) 1849--1851.

\bibitem{BeHa21} E. Bessel-Hagen: \"Uber die Erhaltungss\"atze der Elektrodynamik, {\em Math. Ann.} {\bf 84} (1921) 258--276.

\bibitem{Bau} M. Bauderon: Le probl\`{e}me inverse du calcul des variations,
{\em  Ann. de l'I.H.P.} {\bf 36}(2) (1982) 159--179.

\bibitem{BCDPS13} L. Bonora, M. Cvitan, P. Dominis Prester, S. Pallua, I. Smoli\v c : 
Symmetries and gravitational Chern--Simons Lagrangian terms, {\em  Phys. Lett. } {\bf B 725} (4-5) (2013)  468--472.

\bibitem{BFF03} A. Borowiec,  M. Ferraris,  M. Francaviglia: A covariant formalism for Chern--Simons gravity, {\em  J. Phys. } {\bf A 36} (10) (2003) 2589--2598. 
  
\bibitem{BFFP03} A. Borowiec,  M. Ferraris,  M. Francaviglia, M. Palese: 
Conservation laws for non-global Lagrangians, {\em  Univ. Iagel. Acta Math.} {\bf 41} (2003) 319--331.


\bibitem{BrKr05} J. Brajer\v c \'i k, D. Krupka: Variational principles for locally variational forms, {\em  J. Math. Phys.} {\bf 46} 052903 (2005)

\bibitem{BrKr07} J. Brajer\v c \'i k, D. Krupka:  Cohomology and local variational principles, in  XV International Workshop on Geometry and Physics, {\em  Publ. R. Soc. Mat. Esp.} {\bf 11}  (2007) 119--124.

\bibitem{Cartan22} \' E. Cartan: Lecons sur les invariants int\`{e}graux (Hermann, Paris, 1922).

\bibitem{CISS15} P. Catalan, F. Izaurieta, P. Salgado, S. Salgado:
Topological gravity and Chern--Simons forms in $d = 4$, {\em  Phys. Lett. } {\bf B  751} (2015) 205--208.

\bibitem{CaPaWi16} F. Cattafi, M. Palese, E. Winterroth:  Variational derivatives in locally Lagrangian field theories and Noether--Bessel-Hagen currents, {\em Int. J. Geom. Methods Mod. Phys.} {\bf 13} (8) (2016 ) 1650067.

\bibitem{Ch89}  A.H. Chamseddine: Topological gauge theory of gravity in five and all odd dimensions, {\em Physics Letters B} {\bf 233} (3--4) (1989) 291--294

\bibitem{Ch90}  A.H. Chamseddine: Topological gravity and supergravity in various dimensions, {\em Nucl. Phys. B} {\bf 346} (1) (1990)  213--234.

\bibitem{ChSi71}  S.S. Chern, J. Simons: Some cohomology classes in principal fiber bundles and their application to Riemannian geometry, {\em Proc. Nat. Acad. Sci. U.S.A.} {\bf 68} (1971) 791--794. 

\bibitem{ChSi74} S.S. Chern, J. Simons: Characteristic forms and geometric invariants, {\em Ann. of Math.} {\bf (2) 99} (1974) 48--69. 

\bibitem{Ded55}
P. Dedecker: Quelques applications de la suite spectrale aux int\'egrales multiples du calcul des variations et aux invariants int\'egraux, chapitre I: suites spectrales et g\'en\'eralisations, Bull. Soc. Roy. Sci. Li\`{e}ge 24 (1955) 276--295.

\bibitem{Ded56}
P. Dedecker: Quelques applications de la suite spectrale aux int\'egrales multiples du calcul des variations et
aux invariants int\'egraux, chapitre II: espaces de jets locaux et faisceaux, {\em  Bull. Soc. Roy. Sci. Li\`{e}ge} {\bf 25} (1956) 387--399.

\bibitem{Ded57}
P. Dedecker: Calcul des variations et topologie alg\'ebrique M\'emoires de la Soci\'et\'e
Royale des Sciences de Li\`{e}ge, Quatri\`{e}me s\'erie, Fascicule I, Universit\'e de Li\`{e}ge 19
(1957) 1--215

\bibitem{Ded77}
P. Dedecker: On the generalization of symplectic geometry to multiple integrals in the calculus of variations In: Lecture Notes in Math. 570 (Springer, Berlin, 1977) 395--456.

\bibitem{DT80}
P. Dedecker, W.M. Tulczyjew:  Spectral sequences and the inverse problem of the calculus of variations,
Proc. Internat. Coll. on Diff. Geom. Methods in Math. Phys.,Salamanca 1979, in: {\em Lecture Notes in Math.} {\bf 836} (Springer, Berlin, 1980) 498--503.

%

\bibitem{DJT82} S. Deser, R. Jackiw, S. Templeton: Topologically massive gauge theories, {\em  Ann. Physics}, {\bf 140} (2)  (1982) 372--411.

\bibitem{EdHaTrZa06} J.D. Edelstein, M. Hassa\"{i}ne, R. Troncoso, J. Zanelli: Lie-algebra expansions, Chern--Simons theories and the Einstein-Hilbert Lagrangian, {\em  Phys. Lett.} {\bf B 640} (5-6) (2006) 278--284. 

\bibitem{FeLa98} F. Ferrari, I. Lazzizzera: A Chern--Simons field theory description of topologically linked polymers, {\em Phys. Lett.} {\bf B 444} (1-2) (1998) 167--173.

\bibitem{FeLa99} F. Ferrari, I. Lazzizzera: Topological entanglement of polymers and Chern--Simons field theory, {\em J. Phys.} {\bf A 32} (8) (1999) 1347--1357.

\bibitem{FFPW08}  M. Ferraris, M. Francaviglia, M. Palese, E.  Winterroth: Canonical connections in gauge-natural field theories,  {\em  Int. J. Geom. Methods Mod. Phys.} {\bf }5 (6) (2008) 973--988. 
 
\bibitem{FFPW11}  M. Ferraris, M. Francaviglia, M. Palese, E.  Winterroth: Gauge-natural Noether currents and connection fields, {\em  Int. J. Geom. Methods Mod. Phys.} {\bf 8}(1) (2011)  177--185.

\bibitem{FePaWi10}  M. Ferraris, M. Palese, E. Winterroth: Local variational problems and conservation laws,  {\em Diff. Geom. Appl} {\bf 29} (2011) S80--S85. 

\bibitem{FrPaWi13} M. Francaviglia, M. Palese, E.  Winterroth: Variationally equivalent problems and variations of Noether currents, {\em  Int. J. Geom. Meth. Mod. Phys.} {\bf 10}(1) (2013) art. no. 1220024.

\bibitem{FrPaWi13JPCS} M. Francaviglia, M. Palese, E.  Winterroth:  Cohomological obstructions in locally variational field theories, {\em Jour. Phys. Conf. Series} {\bf 474} (2013) art. no. 012017.

\bibitem{GiMaSa03} G. Giachetta, L. Mangiarotti, G. Sardanashvily: Noether conservation laws in higher-dimensional Chern--Simons theory, {\em Modern Phys. Lett. } {\bf A 18} (37) (2003) 2645--2651. 

\bibitem{Kru90}  D. Krupka: Variational Sequences on Finite Order Jet Spaces, {\em Proc. Diff. Geom. Appl.}; J. Jany\v{s}ka, D. Krupka eds., World Sci. (Singapore, 1990) 236--254.


\bibitem{Kru08}  D. Krupka: Global variational theory in fibred spaces, in {\em Handbook of global analysis}, 773--836, 1215, Elsevier Sci. B. V., Amsterdam, 2008.

\bibitem{Kru15} D. Krupka: {\em Introduction to Global Varational Geometry}, Atlantis Press, 2015.


\bibitem{Lep36} Th.H.J. Lepage: Sur les champ geodesiques du Calcul de Variations, I, II, {\em Bull. Acad. Roy. Belg., Cl. Sci.} {\bf 22} (1936) 716--729, 1036--1046.

\bibitem{Noe18} E. Noether: Invariante Variationsprobleme, {\em Nachr. Ges. Wiss. G\"ott., Math. Phys. Kl.} {\bf II} (1918) 235--257.

\bibitem{PaRoWiMu16} M. Palese, O. Rossi, E. Winterroth, J. Musilov\"a: Variational sequences, representation sequences and applications in physics, {\em SIGMA} {\bf 12 } (2016) 045, 45 pages  (2016).

\bibitem{PaWi02} Palese, M., Winterroth, E.:  Nonlinear $(2+1)$-dimensional field equations from incomplete Lie algebra structures, {\em  Phys. Lett. B} {\bf 532} (1-2) (2002)  129--134.

\bibitem{PaWi07} M. Palese, E. Winterroth:  On the relation between the Jacobi morphism and the Hessian in gauge-natural field theories, {\em Theoret. and Math. Phys.} {\bf 152} (2) (2007) 1191--1200.

\bibitem{PaWi11} M. Palese, E. Winterroth:  A variational perspective on classical Higgs fields in gauge-natural theories, {\em Theoret. and Math. Phys.} {\bf 168} (1) (2011) 1002--1008.

\bibitem{PaWi12} M. Palese, E. Winterroth: Constructing towers with skeletons from open Lie algebras and integrability,
{\em Jour. Phys. Conf. Series} {\bf 343} (2012) 012091.


\bibitem{Tak79} F. Takens: A global version of the inverse
problem of the calculus of variations, {\em J. Diff. Geom.} {\bf 14} (1979) 543--562.


\bibitem{Tul77} W. M. Tulczyjew: The Lagrange complex, {\em Bull. Soc. Math. France} {\bf 105} (1977) 419--431.

\bibitem{Tu80} W.M. Tulczyjew: The Euler--Lagrange resolution, in: Proc. Int. Colloquium on Diff. Geom. Methods in Math. Phys., Aix-en-Provence, 1979, {\em Lecture Notes in Mathematics} {\bf 836} Springer, Berlin, 1980, 22--48.

\bibitem{Tu88} W.M. Tulczyjew: Cohomology of the Lagrange complex, {\em Ann. Scuola Norm. Sup.
Pisa} {\bf 14} (1988) 217--227.

\bibitem{Vin77} A. M. Vinogradov: On the
algebro--geometric foundations of Lagrangian field theory, {\em Soviet Math. Dokl.} {\bf 18} (1977) 1200--1204.


\bibitem{Win16} E. Winterroth: Variational cohomology and global solutions in higher Chern--Simons gauge theory, submitted.

\bibitem{Wi88} E. Witten: $2+1$-dimensional gravity as an exactly soluble system, {\em Nucl. Phys.}  {\bf B 311} (1) (1988) 46--78;

\bibitem{Wi89} E. Witten: Quantum field theory and the Jones polynomial, {\em Commun. Math. Phys.}  {\bf 121} (3) (1989) 351--399.

\end{thebibliography}
\end{document}